\documentclass[aps,prb,groupedaddress,twocolumn,10pt]{revtex4-1}

\begin{document}

\newcommand{\be}{\begin{equation}}
\newcommand{\ee}{\end{equation}}
\newcommand{\bea}{\begin{eqnarray}}
\newcommand{\eea}{\end{eqnarray}}

\title{Demystifying the Holographic Mystique}

\author{D. V. Khveshchenko}

\affiliation{Department of Physics and Astronomy, University of North Carolina, Chapel Hill, NC 27599}

\begin{abstract}

\noindent
Thus far, in spite of many interesting developments, 
the overall progress towards a systematic study and classification of various 'strange' metallic states of matter has 
been rather limited. To that end, it was argued that a recent proliferation of the ideas of 
holographic correspondence originating from string theory might offer a possible way out of the stalemate. 
However, after almost a decade of intensive studies into the proposed extensions of 
the holographic conjecture to a variety of condensed matter problems, the validity of this 
intriguing approach remains largely unknown. This discussion aims at ascertaining its true status 
and elucidating the conditions under which some of its predictions may indeed be right (albeit, possibly, for a wrong reason).
\end{abstract}

\maketitle

{\it {Condensed Matter Holography: The Promise}}\\

Among the outstanding grand problems in condensed matter physics is that 
of a deeper understanding and classification of the so-called 'strange metals' 
or compressible non-Fermi liquid (NFL) states of the strongly interacting systems. 
However, despite all the effort and 
a plethora of the important and non-trivial results obtained with the use of the traditional techniques,  
this program still remains far from completion. 

As an alternate approach, over the past decade there have been numerous  
attempts inspired by the hypothetical idea of holographic correspondence which originated  
from string/gravity/high energy theory (where it is known under the acronym $AdS/CFT$) 
to adapt its main concepts to various condensed matter (or, even more generally, 
quantum many-body) systems at finite densities and temperatures  \cite{hol,books}. 

In its original context, the 'bona fide' holographic principle 
postulates that certain  $d+1$-dimensional ('boundary') quantum field theories
(e.g., the maximally supersymmetric $SU(N)$ gauge theory) may allow for  
a dual description in terms of a string theory which, upon a proper compactification, amounts to a  
certain $d+2$-dimensional ('bulk') supergravity. 
Moreover, in the strong coupling limit (characterized in terms of 
the t'Hooft coupling constant $\lambda=g^2N\gg 1$) and for a large rank $N\gg 1$ of the gauge symmetry group, the bulk 
description can be further reduced down to a weakly fluctuating gravity model which
can even be treated semiclassically at the lowest ($0^{th}$) order of the underlying $1/N$-expansion.

In the practical applications of the holographic conjecture, the partition function of a strongly interacting 
boundary theory with the Lagrangian ${\cal L}(\phi_a)$ would then be approximated by a saddle-point 
(classical) value of the bulk action described by the 
Lagrangian $L(g_{\mu\nu},\dots)$ which includes gravity and other fields dual to their boundary counterparts \cite{hol,books}
\bea
Z[J]=\int\prod_{a=1}^{N}D\phi_a \exp(-\int dtd^{d}{\vec x}{\cal L}(\phi_a))\approx 
\nonumber\\
\exp({-\int drdtd^{d}{\vec x}{L}(g_{\mu\nu},\dots)})
\eea
evaluated with the use of a fixed background metric 
\be
ds^2=g_{tt}dt^2+g_{rr}{dr^2}+\sum_{ij}g_{ij}dx^idx^j
\ee
while any quantum corrections would usually be neglected by invoking the small parameter $1/N$.

Thus, considering that the task of solving a system of coupled Einstein-type differential equations
can be fairly straightforward conceptually (albeit not necessarily technically), 
the holographic approach could indeed become a novel powerful tool for studying 
the strongly correlated systems and a viable alternative to the practically 
impossible problem of summing the entire perturbation series.
Specifically, if proved valid, some of the broad 'bottom-up' generalizations
of the original holographic conjecture known as 'AdS/CMT' (which, in many instances, should have been more appropriately called
'non-AdS/non-CFT') could indeed provide an advanced phenomenological framework for discovering 
new and classifying the already known types of the NFL behavior. 

Thus far, however, a flurry of the traditionally detailed (hence, rarely concise) publications on the topic
have generated not only a good deal of enthusiasm but some reservations as well.    
Indeed, the proposed 'ad hoc' generalizations of the original string-theoretical  
construction involve some of its most radical alterations,  
whereby most of its stringent constraints would have been abandoned   
in the hope of still capturing some key aspects of the underlying correspondence. 
This is because the target (condensed matter) systems generically tend to be neither 
conformally, nor Lorentz (or even translationally and/or rotationally) invariant 
and lack any supersymmetric (or even an ordinary) 
gauge symmetry with some (let alone, large) rank-$N$ non-abelian group.

Moreover, while sporting a truly impressive level of technical profess, the exploratory 
'bottom-up' holographic studies have not yet helped to resolve such crucially important issues as:\\
$\bullet$ Are the conditions of a large $N$, (super)gauge symmetry, 
Lorentz/translational/rotational invariance of the boundary
(quantum) theory indeed necessary for establishing a holographic correspondence with some weakly 
coupled (classical) gravity in the bulk?\\  
\noindent
$\bullet$ 
{Are all the strongly correlated systems (or only a precious few) supposed to have gravity duals?}\\
\noindent
$\bullet$ 
{What are the gravity duals of the already documented NFLs?}\\
\noindent
$\bullet$ 
{Given all the differences between the typical condensed matter and string theory problems, 
what (other than the lack of a better alternative) justifies the adaptation 'ad verbatim' 
of the original (string-theoretical) holographic 'dictionary'?}  
and, most importantly:\\
\noindent
$\bullet$ 
{If the broadly defined holographic conjecture is indeed valid, then {\bf why} is it so?}

Considering that by now the field of CMT holography has grown almost a decade old,
it would seem that answering such outstanding questions should have been considered more important than 
continuing to apply the formal holographic recipes to an ever increasing 
number of model geometries and then seeking some resemblance to the real world systems
without a good understanding as to why it would have to be there in the first place.  
In contrast, the overly pragmatic 'shut up and calculate' approach prioritizes 
computational tractability over physical relevance, thus making it 
more about the method (which readily provides a plethora of answers 
but may struggle to specify the pertinent questions) itself, rather than the underlying physics. 

On the other hand, there exist of course important ongoing efforts towards, both, 
constructing various 'top-down' holographic models \cite{topdown} as well as trying to derive holography 
from the already known concepts such as a renormalization procedure on  
the information-related tensor networks \cite{RG}. 
However, the former approach formulated in terms of such objects as D-branes
remains to be rather exotic and somewhat hard to connect to from the CMT perspective,  
whereas the latter one (which, in practice, amounts to a massive use of the Stratonovich 
and Trotter transformations combined with numerical solutions of the resulting
flow equations of the functional RG-type) has yet to deliver 
a well-defined bulk geometry, other than the basic $AdS$ with the dynamical 
$z=1$ (or its Lifshitz modification with $z=2$), that would be reminiscent of those 
metrics which are extensively utilized in the 'bottom-up' studies (see below). \\

{\it {Condensed Matter Holography: The Evidence}}\\

The circumstantial evidence that would be typically invoked in support of the general idea of
holography includes such diverse topics as thermodynamics of black holes, 
hydrodynamics of quark-gluon plasma and unitary ultracold Fermi-gases,
geometrization of quantum entanglement entropy, etc.
However, while possibly attesting to the validity of some aspects of the holographic concept 
in general, those arguments may not be immediately pertinent to the specific condensed matter systems.

Therefore, for the broadly generalized holographic conjecture to prove relevant and gain a predictive power   
in the latter context its predictions would have to be systematically contrasted against experimental data
as well as the results of other, more conventional, techniques providing a preliminary insight. 
Also, the holographic calculations would have to be carried out for (and allow for a cross-check between) 
a whole range of the thermodynamic and transport quantities, including specific heat, DOS, charge and spin susceptibilities, 
electrical, thermal, and spin conductivities, electron spectral function, etc. 

By some (admittedly, risky) analogy with any 
evidence claimed to support, e.g., the 'science' of UFO and other paranormal phenomena - most 
(but not all) of which can be readily dismissed - in order to 
ascertain the true status of the CMT holography one would need to focus on (and  
identify the physical origin(s) of) those cases of factual agreement 
that can be deemed reliable and reproducible.   
It especially concerns those instances where the holographic predictions were reported to agree quantitatively 
with the results of some exact analytic \cite{Sachdev}
or almost exact (e.g., Monte Carlo) \cite{krempa} calculations (see below). 

At the 'ad hoc' level the holographic approach has already been opportunistically
applied to a great variety of condensed matter systems which 
includes the 'strange' Fermi and Bose metals describing quantum-critical spin liquids, supersolids, 
quantum smectics and nematics,  Mott insulators, (in)coherent conductors, itinerant (anti)ferromagnets, 
Quantum Hall effect, graphene and other Dirac/Weyl metals, multi-channel Kondo and other quantum impurity models, etc. 

Routinely, the holographic calculations would be performed in the 'mean-field' approximation  
(i.e., at the $0^{th}$ order of the would-be $1/N$-expansion) 
and then compared to some selected sets of experimental data on the systems 
that tend to lack any supersymmetry, are characterized by the number of species $N\sim 1$
(such as spin, orbital, and/or valley components) and have only moderate (as opposed to very strong)
interactions. The above caveats notwithstanding, however, such studies would often  
seek nothing short of (and occasionally claim to have found) a quantitative agreement with the data.

Also, many of the early CMT holographic calculations were   
carried out by 'seeking where the lights are' and utilizing just a handful of the historic black-hole solutions,
the central among which is the Reissner-Nordstrom (RN) one that asymptotically approaches the 
$AdS_{d+2}$  (anti-de-Sitter) and $AdS_2\times R^d$ geometries in the UV and IR limits, respectively \cite{hol,books}
\be
g_{tt}= -{f(r)/r^2},~~~ g_{rr}={1/r^2f(r)},~~~
g_{ij}(r)=\delta_{ij}/r^2
\ee
and is accompanied by the scalar potential $A_0=\mu(1-r/r_h)$.
The emblackening factor $f(r)=1-(1+\mu^2)({r/r_h})^{d+1}+\mu^2({r/r_h})^{2d}$ incorporates the 
chemical potential $\mu$ and vanishes at the horizon of radius $1/r_h$ 
proportional to the Hawking temperature $T$ which is shared by the bulk and boundary degrees of freedom.
Notably, this radius remains finite even when the temperature $T=(d+1-(d-1)\mu^2)/4\pi r_h$ vanishes,   
thereby giving rise to the non-vanishing entropy ${\cal S}(T\to 0)\neq 0$ and suggesting 
that the corresponding boundary theory could provide a description of some isolated 'quantum impurity', 
rather than a correlated many-body state with a non-trivial spatial dispersion. 

Accordingly, the boundary fermion propagator $G(\omega,k)$ demonstrates 
the behavior dubbed as 'semi-locally critical'
\cite{semi} which is characterized by a non-trivial frequency, yet a rather mundane momentum, dependence  
\be
G(\omega, {\bf k})={1/(a_{\bf k}+b_{\bf k}\omega^{2\nu_{\bf k}})}
\ee
where $a_{\bf k},~b_{\bf k}$, and $\nu_{\bf k}$ are smooth functions of the momentum $k$.
In the space-time domain, the corresponding behavior 
\be
G(\tau, x)\sim \exp(-S(\tau, x))
\ee
is governed by the (semi)classical action 
$
S(\tau, x)={\sqrt {x^2+(1-2\nu_{\bf k_F})^2\ln^2\tau}}
$
and is consistent with that of the spatially (almost) uncorrelated 'impurities', each 
of which exhibits a characteristic $d=0$ quantum-critical scaling.
Such NFL behavior indeed bears some superficial resemblance to that found in a certain class 
of the heavy-fermion compounds \cite{hf} and DMFT calculations \cite{dmft}. 
However, it is also plagued with such spurious features 
as potentially multiple Fermi surfaces, dispersionless peaks, 
and $log$-oscillating $\omega$-dependence \cite{semi}.

Given the multitude of the experimentally discovered NFLs,
the 'locally-critical' scenario would seem to be much too 
limited to encompass more general types of the real-life NFL scenaria  where, both, 
the $x$- and $\tau$- (or, correspondingly, $\omega$- and $k$-) dependencies of the 
propagator would be distinctly non-trivial.
In light of that realization, the focus of the early holographic studies has gradually
shifted towards a broader class of geometries, including such 
intrinsically non-Lorentz-invariant metrics as the Lifshitz, Shroedinger, helical Bianchi, etc.
A particular attention has been paid to the 
static, diagonal, and isotropic 'hyperscaling-violating' (HV) metrics with the radial 
dependence of the form (up to a conformal equivalence), 
\be
g_{tt}\sim -{r^{2\theta/d-2z}},~~~g_{rr}=g_{ii}\sim {r^{2\theta/d-2}}
\ee  
its finite-$T$ version featuring the additional factor $f=1-(r/r_h)^{d+z-\theta}$.
The dynamical exponent $z$ controls the boundary excitation 
spectrum $\omega\propto q^z$, while $\theta$ quantifies  
a non-trivial scaling of the interval $ds\to \lambda^{\theta/d} ds$,
the scaling-(albeit not Lorentz-) invariant case of the Lifshitz metric corresponding to $\theta=0$.
The physically sensible values of $z$ and $\theta$ must satisfy the 'null energy conditions' 
\be
(d-\theta)(d(z-1)-\theta)\geq 0,~~
(z-1)(d+z-\theta)\geq 0
\ee
signifying a thermodynamic stability of the corresponding geometry.

A proper choice of $\theta$ determining the 'effective dimension' $d_{eff}=d-\theta$ was discussed in the context of fermionic entanglement entropy which singles out the value $\theta=d-1$, consistent with the notion of the Fermi 
surface as a $d-1$-dimensional membrane in the reciprocal (momentum) space \cite{huijse}. 

The HV metrics arise among the solutions of various generalized gravity theories, including those with massive vector (Proca) fields, Horava gravity, as well as the Einstein-Maxwell-dilaton (EMD) 
theory which includes an additional neutral scalar ('dilaton') field \cite{hv}, 
\be
L_{EMD}={1\over 2}(R+{d(d+1)\over L^2})-{(\partial\phi)^2\over 2}-U(\phi)
-{V(\phi)\over 2}F_{\mu\nu}^2
\ee,
where the dilaton potential $U(\phi)$ and the effective gauge coupling $V(\phi)$ are given by
some (typically, exponential) functions of $\phi$.  
This model is believed to be dual to a strongly interacting 
boundary theory at finite density and temperature which is deformed away from the 
hyperscaling limit by a relevant neutral scalar operator dual to the dilaton.  
The HV solutions of the coupled Einstein-Maxwell equations have also been obtained by 
taking into account a back-reaction of the fermionic matter fields on the background geometry \cite{electronstar}.
It should be noted, though, that apart from a few exceptions \cite{Sachdev1}  
such analyses were limited to the hydrodynamic (Thomas-Fermi) description  
of the fermions, while leaving out their more subtle (exchange and correlation) effects.

One purported success of the theory (8) (with $U(\phi)\sim \phi^2$ and $V(\phi)=const$) was  
a numerical fit to the experimentally observed power-law behavior of the mid-infrared optical conductivity \cite{2/3}
\be
\sigma(\omega)\sim\omega^{-2/3}
\ee
in the normal state of the superconducting 
cuprates such as $BSCYCO$ \cite{marel}.
Notably, though, such an agreement was found over less than half of a decade ($2<\omega\tau<8$) 
while the later studies did not confirm it \cite{no2/3}. Besides, 
it was also argued to be intermittent with the alternate  
'universal', $\sim\omega^{-1}$, behavior. 

Another notable example is the holographic calculation \cite{homes2}
of the numerical prefactor $C$ in the empirical Homes relation  
between the superfluid density, critical temperature, and normal state conductivity \cite{homes1}
\be
\rho_s=CT_c\sigma(T_c^{0+})
\ee
This relation is obeyed by a large variety of 
superconductors, including the weakly coupled (yet, sufficiently disordered) 
conventional ones where a justification for the use of 
holography would be rather hard to come by and where the Homes' relation can be shown to hold
within the scope of the traditional theory \cite{kogan}. 
The holographic theory of Ref.\cite{homes2} claimed its success with finding the prefactor $C\approx 6.2$ to be close  
to the experimental values  in the cuprates  
which happen to be $C\approx 8.1$ and $4.4$ for the $ab$- and $c$-axes transport, respectively.
Somewhat ironically, though, this would seem like a case where,
in the absence of a compelling underlying reason, 
a perfect quantitative agreement could do more harm than good to the cause. 

Arguably, a stronger case could have been made if the range of the possible 
values of $C$ were unusually narrow, thus potentially hinting at some universality of the result. 
This, however, does not seem to be the case.
Moreover, the holographic calculation of Ref.\cite{homes2}
does not readily reproduce 
any other empirical relations such as the Schneider's one \cite{schneider} 
\be
{T_c\over T_c^*}=2{\sqrt {\rho_s\over \rho_s^*}}(1-{1\over 2}{\sqrt {\rho_s\over \rho_s^*}}),
\ee
thereby suggesting that the reported agreement with experiment may have 
been largely fortuitous. 

Also, to further strengthen the case for holography, some works tend to use 
increasingly more and more complicated models with a larger number of 
fields while operating under the assumption
that the IR physics should be fairly universal and insensitive to such details. 
The results, on the contrary, indicate that there is little universality, 
as by varying the contents of the bulk theory one
can dramatically alter the boundary behavior - e.g., reproducing on demand     
the entire phase diagram of the cuprates with all the four main phases as well as the domains of their coexistence 
in the extended EM model with the additional vector and scalar fields which represent the competing orders \cite{kiritsis_last}.

There has also been some effort on the experimental side, such as
the report of measuring the shear viscosity-to-entropy density ratio 
in the cuprates \cite{bnl} which was found to be close to its
celebrated holographic lower bound value \cite{kss}
\be 
\eta/{\cal S}=1/4\pi
\ee 
It should be noted, though, that instead of such genuine two-particle characteristic 
as the correlation function of the stress tensor's component $T_{xy}$
in Ref.\cite{bnl} the viscosity $\eta$ was deduced from the single-particle 
electron spectral function. 
Furthermore, even the bound (12) itself is known not to be totally universal,
its value becoming lower, e.g., in the absence of the spatial rotational invariance  
\cite{anisotropy}
\be
\eta_{xzxz}/{\cal S}=(1/4\pi)(g_{xx}/g_{zz})
\ee
Yet another example of a successful application of the holographic approach  
was presented in Ref.\cite{krempa} where an impressive agreement was found between 
the holographically computed frequency-dependent (finite-temperature) 
conductivity $\sigma(\omega)$ and the Monte Carlo results for the Bose-Hubbard model
whose critical behavior belongs to the universality class of the $O(2)$-symmetrical 
Wilson-Fisher critical point. Upon analytically continuing
to the real frequencies this agreement gets progressively worse for $\omega<2\pi T$, though \cite{krempa_last}.
Most intriguingly, the original work of Ref.\cite{krempa} utilized the EM
holographic model with a rather special higher-derivative term
\be
\Delta L = C_{abcd}F^{ab}F^{cd}
\ee
where $C_{abcd}$ is the Weyl tensor, thus making one wonder as to the 
reasons behind a seemingly unique role of this model as the potential Bose-Hubbard's dual. 
However, in the later Ref.\cite{krempa_last} nearly identical results were obtained 
with the use of a much less exotic EMD model with $U(\phi)=\phi^2$ and $V(\phi)=1+\alpha\phi$. 
Thus, the previously reported agreement with the MC results (limited to
$\omega>2\pi T$) appears to be rather common, which observation takes away much of the intrigue
surrounding the holographic model equipped with the term (14) and makes 
less pressing the need for understanding its otherwise inexplicably serendipitous success. 

In light of the above, one would be led to the conclusion that at its current stage  
the CMT holography still lacks a true 'smoking gun' that could lend a firm support for this approach 
(albeit, possibly, in some reduced, rather than its most broad, overreaching, form).
Therefore, it is perhaps not too surprising that there is still no consensus, neither 
on the exact implications of the reports of some apparent agreement between
the holographic predictions and the results of other approaches and/or experimental data,  
nor the general applicability and the principal limitations of the holographic approach itself. \\ 

{\it {Emergent Geometry and 'Holography Light'}}\\

Regardless of those practices in the field of CMT holography that make it prone 
to criticism, its general idea is undeniably appealing. 
Indeed, it falls very much in line with some of the most profound paradigms in 
quantum many-body physics, including those of the bulk-edge correspondence and the notion of 
running couplings in the process of renormalization
alongside the changing scale of energy/length/information. The former has recently received a lot of attention with the advent of topological insulators/superconductors and Dirac/Weyl (semi)metals, while the latter provides a standard framework for relating the bare (UV) to the effective (IR) physics.

Arguably, of all the holographic studies the most important is the quest into its possible underlying physical 
cause(s).  This concerns, first and foremost, those situations where some precise - analytic or numerical - agreement has been found between the results of the holographic and some conventional approaches. 
One such recent example is provided by the random infinite-range fermion-hopping models  
that were studied in Ref.\cite{Sachdev}, drawing from the previous analyses of the 
random infinite-range spin-coupling systems \cite{Sachdev_Ye}.

These models also share a number of common features with the Kondo lattice \cite{Kondo_French} and matrix \cite{matrix} ones,  
their unifying physical theme being that of a single 
'quantum impurity' interacting self-consistently with a local bath.  
This behavior makes them potentially amenable to the description based on the 
holographic 'semi-locally critical' scenario of Refs.\cite{semi}.

Indeed, in Refs.\cite{Sachdev} an emergent invariance under the group of reparametrizations 
was discovered and its relation to 
the 'semi-local' $AdS_2\times R^d$ geometry was demonstrated by establishing 
an exact agreement between the holographically computed single-particle propagators in this (non-fluctuating) geometry \cite{semi} and the two- and four-point correlation functions of certain exactly 
solvable random fermion-hopping models. 

Notably, such a perfect agreement was achieved without invoking a small $1/N$ parameter, 
thereby suggesting that the $1/N$-corrections (that would have been present in the 
original string-theoretical $AdS/CFT$ correspondence) may, in fact, be absent altogether. 
More precisely, although the large-$N$ limit indeed had to be taken in the random spin-coupling model of Ref.\cite{Sachdev_Ye}
(after taking the limit of a large number $\cal N$ of the lattice sites ),
the 'semi-local' behavior of its random fermion-hopping counterparts 
sets in already at $N=1$ (but still for $\cal N>>1$) \cite{Sachdev}. 

Generalizing the results of Refs.\cite{Sachdev}, 
one can construct a whole family of 'semi-local' NFL regimes described by the 
propagator $G(\tau)$ which obeys the integral equations ($n=2$ in Refs.\cite{Sachdev})
\bea
\int G(\tau_1,\tau)\Sigma(\tau,\tau_2)d\tau=\delta(\tau_1-\tau_2),
\nonumber\\
\Sigma(\tau_1,\tau_2)=\lambda G^{n}(\tau_1,\tau_2)G^{n-1}(\tau_2,\tau_1)
\eea
that are manifestly invariant  
under an arbitrary change of variables $\tau=f(\sigma)$
\be
G(\tau_1,\tau_2)=[f^{\prime}(\sigma_1)f^{\prime}(\sigma_2)]^{-1/2n}
G(\sigma_1,\sigma_2)
\ee
To derive these equations from the (disorder-averaged) Hamiltonian 
of a random infinite-range fermion-hopping model
\be
{H}=\sum_{k=1}^n c_k \sum_{a_i,b_i}\psi^\dagger_{a_1}\dots\psi^\dagger_{a_k}\psi_{b_1}\dots\psi_{b_k}
\ee
where the parameters  $c_k$ need to be fine-tuned
\cite{future}, while for the their arbitrary values the system conforms to 
the generic behavior described by the $n=2$ case.  
This is somewhat reminiscent of the situation with the exactly soluble $1d$ spin-$S$ chain model whose 
behavior is markedly non-generic for all $S>1/2$ \cite{spinchain}. 

Specifically, the solution of Egs.(15) demonstrates an expressly NFL behavior 
\be
G(\tau)\sim\tau^{-1/n}
\ee
or, equivalently, $\Sigma(\omega)\sim \omega^{1-1/n}$
which can also be reproduced holographically with the use of a massive bulk fermion 
subject to the $AdS_2\times R^d$ geometry and described by the action 
$
L_f=i{\bar \psi}\gamma^{\mu}\partial_{\mu}\psi-m{\bar \psi}{\psi}
$
provided that the fermion mass $m$ is chosen as follows  
\be
\Delta_{\psi}={1\over 2}-(m^2R^2-q^2{\cal E}^2)^{1/2}=1/2n 
\ee
where $R$ is the $AdS$ radius, $q$ is the fermion charge, and
$\cal E$ measures the particle-hole asymmetry \cite{Sachdev}.
Thus, regardless of whether or not the 'semi-locally critical' scenario of Ref.\cite{semi}  
is immediately applicable to any real-life materials, 
its predictions might still prove quite useful, as far as the properties of some properly crafted 
random spin-coupling and fermion-hopping models are concerned.

Notably, though,  
the dimension (19) arises in the so-called 'unstable' boundary CFT which is dual to $AdS_2$
\cite{semi}, while in the 'stable' one (to which the former is supposed to flow under a double-trace deformation) 
the operator dimension would be $\Delta_+={1\over 2}+(m^2R^2-q^2{\cal E}^2)^{1/2}>1/2$, thus making the value 
(19) unattainable for any integer $n>1$. 

It is also worth noting that the behavior similar to Eq.(18) 
can be envisioned in the context of, e.g., sub-Ohmic spin-bath models
which, incidentally, can be formulated in terms of the localized Majorana fermions.
Among other things, it gives rise to the ubiquitous Lorentzian ('Drude-like') 
frequency dependencies of the various observables, their width 
being given by the inelastic phase relaxation rate \cite{shnirman}. 
It might be interesting to explore such a possible connection further \cite{future}, as well as to reproduce
Eq.(18) in the holographic models aspiring to describe the Kondo physics \cite{hol_Kondo}. 

Also, while being characteristic of the purely classical (non-fluctuating) 
asymptotic near-horizon geometry \cite{semi} 
the emergent reparametrization symmetry (15) does not rise to the same level as that in the elaborate  
constructs of the original (string-theoretical) holographic correspondence
where the bulk supports  a full-fledged quantum (super)gravity theory that only becomes classical 
in the large-$N$ limit \cite{hol}.
Nevertheless, it is quite remarkable that the quantum systems in question 
allow for some of their properties to be expressed in purely geometrical terms.

In that regard, emergent classical metrics and concomitant effective gravity-like descriptions 
are not that uncommon, the most remarkable example being provided by the intriguing relationship   
between quantum physics and classical geometry in the form of 
the famous extremal area law of entanglement entropy \cite{ryu}. 
To a somewhat lesser extent, same can be said about such connections found in
thermodynamics of phase transitions ('Fisher-Ruppeiner metric') \cite{thermometric},
quantum information and tensor network states \cite{mera},
QHE hydrodynamics ('Fubini-Study metric') \cite{fubini}, 
Chern classes of the Bloch eigenstates of momentum \cite{bloch}, 
Berry phase of the adiabatic time evolution \cite{evo}, etc.  

Importantly, the emergent geometric structures such as effective metrics, curvatures, spin connections, etc.
can indeed create the appearance of a limited form of some bulk-boundary correspondence ('holography light') 
that can be inadvertently mistaken for manifestations of the hypothetical full-fledged CMT holography.\\

{\it {Holographic Phenomenology: The Cuprates}}\\

Obtaining verifiable holographic predictions can be particularly instructive in those situations 
where experimental data exhibit robust scaling dependencies, 
thus hinting at a possible (near-)critical regime
that might be amenable to a holographic scaling analysis \cite{conductivity}. 
One such example is provided by the holography-inspired phenomenologies of the cuprates which
focused on the robust power-law behaviors of the longitudinal electrical conductivity, Hall angle,
and magnetoresistivity that violates the conventional Kohler's law  \cite{kohler}
\bea
\sigma\sim T^{-1}\nonumber\\
\tan\theta_H\sim T^{-2},\nonumber\\
\Delta\rho/\rho\sim \rho^2
\eea
In the previously proposed scenaria, Eqs.(20) were argued to indicate a possible existence of two distinctly different  
scattering times: $\tau_{tr}\sim 1/T$ and  $\tau_{H}\sim 1/T^2$ which characterize the  
longitudinal vs transverse \cite{anderson}
or charge-symmetric vs anti-symmetric \cite{coleman} currents.
Yet another insightful proposal was put forward in the framework of the marginal Fermi liquid phenomenology \cite{varma}. 

In contrast to the anomalous transport properties of the cuprates,
their thermodynamic ones are more standard, including the Fermi-liquid-like specific heat 
$C\sim T$ (except for a possible logarithmic enhancement \cite{specificheat}).
A recent attempt to rationalize such experimental findings
was made in Refs.\cite{hk,pheno}. Under the assumption of an underlying one-parameter scaling 
the dimensions of the observables were expressed in terms of the minimal set
which includes the dynamical critical exponent $z$ alongside 
the dimensions of the (mass) density $\Delta_n$ and  
effective charge $\Delta_e$, the two accounting for, roughly speaking,  
the wave function renormalization and vertex corrections, respectively.

The linear thermoelectric response 
is then described in terms of a trio of the fundamental kinetic coefficients
\bea
{\bf J}={\hat \sigma} {\bf E}-{\hat \alpha} {\bf \nabla}T
\nonumber\\
{\bf Q}=T{\hat \alpha} {\bf E}-{\hat {\kappa}} {\bf \nabla}T
\eea
where the Onsager's symmetry is taken into account
and the off-diagonal entries in the $2\times 2$ matrices ${\hat \sigma}, {\hat \alpha}, {\hat {\kappa}}$ 
represent the Hall components of the corresponding conductivities. 

To set up the scaling relations that reproduce Eqs.(20) and other experimentally observed algebraic 
dependencies one has to properly account for the time reversal and particle-hole symmetries.
Also, one should be alerted to the fact that the exponents governing the temperature dependencies in the kinetic coefficients 
would be the same as those appearing  
in their frequency-dependent optical counterparts \cite{conductivity}.
However, the previous analyses based on the semiclassical kinetic equation show that such 
leading (minimal) powers of $T$ may or may not actually survive, depending on whether or not 
the quasiparticle dispersion and Fermi surface topology 
conspire to yield comparable rates of the normal and 
umklapp inelastic scattering processes \cite{maslov}.  
It would appear, though, that in the cuprates, both, the multi-pocketed (in the under- and optimally-doped
cases) as well as the extended concave (in the over-doped case) hole Fermi surfaces 
comply with such necessary conditions.

As an additional consistency check, the kinetic coefficients 
are expected to be consistent with the Fermi liquid relations
$
S\sim (T/e\sigma){d\sigma/d\mu}, ~~~\nu_N\sim (T/eB){d\theta_H/d\mu} 
$
where the r.h.s. are proportional to the Fermi surface curvature
but do not contain such single-particle characteristics as scattering time or effective mass
and, therefore, might be applicable beyond the coherent quasiparticle regime. 

It was found in \cite{pheno} that, somewhat surprisingly, most of the experimental data 
favor the rather mundane solution  
\be
z=1,~~\Delta_e=0,~~\Delta_n=1
\ee
which imposes the following relation on 
the Seebeck ($S$), Hall Lorentz ($L_H$), and Nernst ($\nu_N$) coefficients 
(barring any Sondheimer-type cancellations)  
\be
[S]+[\nu_N]-[L_H]=0
\ee
In the optimally doped cuprates, the experimentally measured thermopower, apart from a finite offset term, demonstrates a (negative) linear $T$-dependence \cite{thermopower}.
As regards the Hall Lorentz and Nernst coefficients, the data on the untwinned samples of
the optimally doped $YBaCuO$ were fitted into a 
linear dependence, $L_H\sim T$, whereas $\nu_N$ was generally 
found to decrease with increasing $T$, thus suggesting $[\nu_N]<0$ \cite{ong}. 
Moreover, the Nernst signal increases dramatically with decreasing temperature, which effect 
has been attributed to the superconducting fluctuations and/or fluctuating vortex pairs whose (positive) contribution dominates over the quasiparticle one 
(the latter can be of either sign, depending on the dominant type of carriers) upon approaching $T_c$.
Besides, $\nu_N$ turns out to be strongly affected by a proximity to the 
pseudogap regime and can even become anisotropic.

The 'Fermi-surfaced' solution (22) should be compared with
the significantly more exotic one obtained in Ref.\cite{hk} 
\be
z=4/3,~~\Delta_e=-2/3,~~\Delta_n=2
\ee
where the thermoelectric coefficients
$
S\sim T^{1/2},~~L_H\sim T,~~\nu_N\sim T^{-3/2}
$
do not obey Eq.(23). Also, in this case 
an agreement with the linear behavior of $L_H$
reported in Ref.\cite{ong} is guaranteed
by imposing it as one of the constitutive relations.
In spite of this predestined success, though, this scheme  
fails to reproduce the linear (up to a constant) thermopower, although it claimed 
the alternate $S=a-bT^{1/2}$ dependence to provide an even better fit to 
the data \cite{thermopower}.  

Also, the solution (24) predicts $C\sim T^{3/2}$
and, therefore, appears to be at obvious odds with the observed thermodynamic properties as well \cite{specificheat}.
Moreover, it predicts the linear in $T$ (longitudinal) Lorentz ratio 
$
L={\kappa/T\sigma}
$
similar to its Hall counterpart, contrary to 
a constant $L$, as suggested by the scenario (22). 

Although an additional experimental effort is clearly called for in order to discriminate more definitively 
between the above predictions, it should be mentioned that more recently a slower-than-linear 
temperature behavior of the (electronic) Lorentz ratio has been reported \cite{matusiak}.

Lastly, the solution (24) features $\theta=0$ and strong (infrared) 
charge renormalization ($\Delta_e<0$). 
Since in the presence of a well-defined $d$-dimensional Fermi surface the 'hyperscaling-violation' parameter 
is expected to coincide with its co-dimension, hence $\theta=d-1$ \cite{huijse}, 
the value $\theta=0$ hints at a point-like ('Dirac') Fermi surface (if any).
In contrast, Eq.(22) suggests a 
rather simple physical picture where neither the Fermi liquid-like dispersion, nor 
the effective charge demonstrate any significant renormalization. 
Thus, to construct a viable phenomenological description of the optimally doped cuprates one might be able to do away without introducing the additional charge exponent $\Delta_e$,  
contrary to the assertions made in Refs.\cite{hk,guteraux}.

It might also be worth mentioning that 
another long-time baffling, the underdoped, phase of the cuprates has recently been downgraded 
to a potentially simpler organized state of matter with some of its properties being Fermi-like \cite{mundane}.   

Also, the recent attempt to match the exponents (24) with the predictions of the 
semi-phenomenological 'unparticles' model showed that it can only be 
achieved at the expense of a further, multi-flavored, extension 
of that scenario  \cite{phillips}. However, the sheer complexity of the proposed construction seems to 
indicate rather clearly that the latter can hardly provide a natural (let alone, minimal) 
theory of the anomalous transport in the cuprates.  

Moreover, in the last of Refs.\cite{phillips}
the method of fractional derivatives was employed to generalize the Maxwell's and continuity equations,
\bea
\partial^{\alpha}_{\mu}F^{\mu\nu}=J^{\nu}\nonumber\\
\partial^{\alpha}_{i}J^{i}+\partial_0\rho=0
\eea
However, it can be readily seen that Eqs.(25) are only consistent when the dimension of the 
vector potential $[A_i]=[\partial^{\alpha}_i]=\alpha$ equals $z$ (which was chosen as $z=1$ in Ref.\cite{phillips} ), 
thus prohibiting it from taking any anomalous value. 

In fact, a gradual realization of how strained can be the attempts to squeeze the holographic phenomenology of the cuprates
into the 'Procrustes bed' of the two-parametric HV geometries may have already started making its way into   
the holographic community \cite{baggioli}.

Another viable candidate for applications of the holographically-inspired scaling theory
is provided by the self-consistent strong-coupling solutions aimed at 
describing the properties of certain $2d$ and $3d$ antiferromagnetic metals.
Analyzed under the assumption of
a momentum-independent NFL self-energy, $\Sigma(\omega) \sim\omega^{1-\alpha}$,
it yields the critical exponents as sole functions of the spatial dimension \cite{woelfle}
\bea
\alpha=1/2-1/z_b,~~~z_b=4d/3,~z_f={1/(1-\alpha)},\nonumber\\
\nu={1/(2+z_b\alpha)}
\eea 
Notably, the hyperscaling relations are still obeyed, resulting in the anomalous dependencies 
\be
C\sim T^{1-\alpha},~\sigma\sim\omega^{\alpha-1},~\chi_s\sim T^{\alpha-1}
\ee
In the presence of disorder these results are likely to be modified, though \cite{belitz}.

Another recent work \cite{maier_strack} utilized the functional RG technique, thus obtaining a different solution
which might be pertinent to the AFM metals: $z_f=z_b=3/2,~\theta=0,~\nu=\gamma=1$.  
Reproducing these solutions and searching for the new ones \cite{future} poses an interesting challenge and presents a
important test for the holographically inspired scaling theory of these materials.\\

{\it {From AdS/CMT to Holographic Transport}}\\

A great many of the early works on CMT holography   
utilized the standard holographic recipe for
computing electrical conductivity and other kinetic coefficients as boundary limits of the ratios
 between the convergent (associated with a response) and divergent (associated with a source) terms in the 
 solutions of the radial classical equations \cite{hol}. 
This prescription would result in, e.g., the expression for the electrical conductivity  
\be
\sigma(\omega)
={1\over i\omega}{B\over A}={1\over i\omega}r^{2-d}{d\ln F_{ri}\over dr}|_{r\to 0}
\ee
where the component of the field tensor $F_{ri}=A+Br^{d-1}/(d-1)$ is a bulk dual of the boundary electric current.

Moreover, in the absence of any dilaton accounting for the scale dependent couplings 
and bringing about a dependence on the radial holographic coordinate, 
the physical observables become anomaly-free and scale-invariant, hence unaffected by renormalization.
In that case they can be cast in a purely algebraic form in terms of the horizon metric,
as per the 'membrane paradigm' \cite{jain}. 
For instance, the electrical conductivity and charge susceptibility read 
\bea
\chi=(\int^{r_h}_{0} dr{g_{rr}g_{tt}\over (-g)^{1/2}})^{-1} \nonumber\\
\sigma= ({-g\over g_{rr}g_{tt}})^{1/2} g^{ii}|_{r_h},
\eea
where $g$ is the determinant of the metric. Per the Einstein's relation
he two are related via the diffusion coefficient, $\sigma=\chi_cD$.

However, while it might be possible to predict such general features as the exponents 
in the power-law frequency dependences of the optical electrical and thermal conductivities \cite{conductivity}
by using the formal prescriptions similar to  Eq.(28,29), the latter  
do not properly account for the actual physical contents of the boundary theory and a potentially intricate interplay 
between its different scattering mechanisms.

A list of the proposed sources of current and/or momentum relaxation (which is often  
confusingly referred to as 'dissipation') employed in holography
includes random boundary and horizon potentials, helical (Bianchi $VII_0$-type) 
and spatially periodic ('Q-lattice') geometries, massive gravity, axion fields, etc. \cite{transport}.  

Common to all the different approaches, though, 
is the unifying fact that the conductivity and other kinetic coefficients would be typically 
given by a sum of two terms
\be
\sigma(\omega)=\sigma_0 + {e^2n^2\over \chi_{PP}(\gamma-i\omega)}
\ee
although in the 'top-down' constructions using the DBI action  \cite{topdown} Eq.(30) would be replaced by 
$\sigma={\sqrt {\sigma^2_0+\sigma^2_1}}$. 

In Eq.(30)  the momentum susceptibility $\chi_{PP}=E+P$ is equal to the enthalpy density.
and the first term corresponds to the (potentially, universal) diffusion-limited contribution to the conductivity 
that survives in the particle-hole symmetric limit of zero charge density ($n\to 0$). 
It is often associated with the processes of 'pair creation' by which
a neutral system develops a non-vanishing (yet, finite) conductivity in the absence of momentum relaxation ($\gamma\to 0$), while
the DC ($\omega\to 0$) conductivity of a finite density system is 
governed by the second term in (30), becoming infinite in the limit $\gamma\to 0$.  

Moreover, in the case of Q-lattices and $1d$ periodic potentials (but not those that break
translational invariance in all the spatial dimensions) the conductivity can still be expressed in terms of 
the horizon data, thus generalizing the case of zero charge density and no momentum relaxation.
Interestingly, though, even for generic ('non-Q') lattices which break translational symmetry in all
the spatial dimensions the conductivity can be obtained by solving some linearized, time-independent, and forced
Navier-Stokes-type equations of an effective (charged and incompressible) fluid living on the horizon \cite{NS},
thus extending the 'membrane paradigm' along the lines of the general concept of a fluid-gravity correspondence \cite{fluid_grav}.
The latter allows for a dual description of the bulk gravity theory in terms of the hydrodynamics of
a certain boundary fluid whose stress-energy tensor acts as a source for the boundary metric.
However, while unveiling yet another intrinsically geometric aspect
of an underlying quantum dynamics, the fluid-gravity correspondence is clearly not 
identical to (albeit, possibly, far more general than) the original string-theoretical holography. 

In certain axion models \cite{davison}, the elastic rate can be readily computed 
although the resulting behavior $\gamma\sim max[T^2,m^2]/T$ can hardly remain physical in the $T\to 0$ limit.
Remarkably, though, at $m={\sqrt 2}r_h={\sqrt 8}\pi T$ the linearized (classical  
gravitational) equations of motion appear to be exactly solvable, yielding the frequency- and momentum-dependent
response functions $G(\omega,k)$ in a closed form and also signaling an emergent 
$SL(2,R)\times SL(2,R)$ symmetry  \cite{davison}.
 
In the more general situation, 
momentum conservation would be broken by some operator $\cal O$ with the dimension $\Delta_O$ and the corresponding 
elastic scattering rate $\gamma$ can be expressed in terms of the spectral density 
$D(\omega,k)=Im<{\cal O}(\omega,k){\cal O}(-\omega,-k)>$ \cite{transport}
\be
\gamma={1\over d\chi_{PP}}\int {\bf dk} k^2{Im D(\omega,k)\over \omega}|_{\omega\to 0}
\ee  
In the EMD context, the following non-universal behavior 
was obtained by treating the pertinent 'random Ising magnetic field' disorder perturbatively  \cite{patel},  
\be
\sigma(T)\sim T^{2(z-1-\Delta_O)/z}
\ee 
Also, the generalized Harris criterion for the relevance of disorder  
\be
\Delta<d+z-\theta/2 ~~~\Delta<(d-\theta)/2+z
\ee
gets saturated for $\Delta_O=(d-\theta)/2+z$, resulting in the universal, yet $\theta$-dependent, 
behavior \cite{patel}
\be
\sigma\sim T^{(\theta-2-d)/z}\sim \sim 1/({\cal S} m^2)
\ee 
where the concomitant scaling of the entropy density, 
$
{\cal S}\sim T^{(d-\theta)/z}
$,
has been taken into account.
In Eq.(34) the second factor can be interpreted as the 'thermal graviton mass', $m\sim T^{1/z}$.
Thus, it is only in the extreme 'locally-critical' limit  $z\to\infty$
where the famous prediction $\sigma\sim 1/S$ of Ref.\cite{zaanen} could be confirmed 
without any sleight of hands.

Moreover, should the relevant bosonic modes
happen to be transverse and, therefore, protected from developing a mass on the grounds of unbroken gauge invariance 
(as it would be the case in any $SU(2)$- or $U(1)$-symmetric spin liquid state), 
the problem of computing the conductivity would require a full non-perturbative solution.

Notably, even in the HV model with $z=3/2$ and $\theta=1$ (which values satisfy the relation 
$z=1+\theta/d$ saturating the second of Eqs.(7))
proposed as a viable candidate for the bulk dual of the theory of
$2d$ fermionic matter coupled to an overdampted bosonic (e.g., gauge) 
field  \cite{huijse,dvk} the conductivity exponent in Eq.(34)
($(\theta-2-d)/z=-2$) would be different from its target value ($-1$).
This observation, too, might be pointing at some deeper problem with reproducing the linear normal state resistivity 
of the cuprates in the fremework of the HV holographic models.

More recently, the main focus of the CMT holographic phenomenology began to shift towards 
developing a general 'holography-augmented' transport theory \cite{davison,hydro,lucas}. This commendable 
effort strives to establish general relations between the transport coefficients 
and find their bounds (if any) that could remain valid regardless of the (in)applicability of 
the generalized holographic conjecture itself.
Such relations are expected to hold in the hydrodynamic regime
governed by strong inelastic interactions
where the rates of momentum relaxation due to elastic disorder, lattice-assisted inelastic (Umklapp)  
electron-electron and electron-phonon (outside of the phonon drag regime) scattering  
are all much smaller than the universal inelastic rate $\Gamma\sim T$ of the normal collisions controlling 
the formation of a thermalized hydrodynamic state itself \cite{kss,zaanen}. 

Notably, the onset of hydrodynamics is a distinct property of the strongly correlated systems, 
as it is unattainable in the standard FL regime where all the 
local (in the $k$-space) quasiparticle densities $n_k$ are nearly conserved.
The hydrodynamic regime does not set in for $d=1$ either due to the peculiar $1d$ kinematics which facilitates  
the emergence of infinitely many (almost) conserved densities. 

In the hydrodynamic regime, the holographic results have also been systematically  
compared to those of the hydrodynamic \cite{hydro}  and memory matrix \cite{lucas} formalisms
which do not rely on the existence of well-defined quasiparticles.
The kinetic coefficients obtained by means of these alternate techniques  
appear to be similar to the holographic expressions (30),  
identifying the density $n$ with the current-momentum susceptibility  $\chi_{JP}=<J|P>=n$ 
which quantifies the contribution to the current from 'momentum drag'. 
Such formalism provides an elegant and physically transparent 
language for discussing, e.g., a decoupling of the
electric current onto the coherent component parallel (in the functional sense) to the 
momentum $P$ and the orthogonal, incoherent, one which has no overlap with $P$ and is responsible
for the universal conductivity of the neutral (particle-hole symmetric)  system 
\be 
J={\chi_{JP}\over \chi_{PP}}P+J_{incoh}
\ee
Although under a closer inspection the hydrodynamic results were found to be 
subtly different (beyond the leading order) from the earlier holographic ones,
such differences have been reconciled by including the corrections to $\sigma_o$ and the residue of the pole in Eq.(30)
at $\omega=-i\gamma$ ('Drude weight') \cite{davison,hydro}. 
This agreement between the holographic and 
hydrodynamic/memory matrix analyses 
was argued to provide an additional evidence supporting (at least, the transport-related aspects of) the former. 

It should be noted, though, that the AC kinetic coefficients 
akin to Eq.(30) appear to have a strictly monotonic frequency dependence 
which is governed by the coherent zero-frequency Drude peak whose 
presence reflects the existence of a (nearly) conserved momentum which overlaps with the electrical current for $n\neq 0$.
 
On the other hand, a non-monotonic - or, else, strictly universal (frequency-independent) - 
behavior would be necessary in order for the conductivity to comply   
with the holographic sum rules proposed in Refs.\cite{krempa}:
\bea
\int^{\infty}_0(\sigma(\omega)-\sigma(\infty))d\omega=0
\nonumber\\
\int^{\infty}_0({1\over \sigma(\omega)}-{1\over \sigma(\infty)})d\omega=0
\eea
These sum rules account for a transfer of the spectral weight
from the coherent Drude peak to the incoherent high-frequency tail. 
Such a non-monotonic behavior develops at $\omega\sim \Gamma$ and may or may not be detectable by the hydrodynamic/memory matrix 
analyses, though.  

One of the most striking predictions of the 
memory matrix calculations is that of the reported absence of (many-body) localization 
and diffusion-dominated metallic transport in strongly interacting 'strange metals' \cite{noMIT}.
It would seem rather surprising, though, that while imposing robust conductivity bounds in 
$d=1$ and $2$ (where in the latter case the lower bound even seems comparable to the 
upper one), this approach fails to find such bounds in $d>2$
where, according to the general wisdom, localization would be even less likely to set in. 

At a deeper level, a fully quantum description of localization (or a lack thereof) still 
seems to remain out of reach in all the existing holographic treatments of disorder. 
In fact, any 'mean-field' approximation is likely to be in principle incapable of taking into account not only the 
localization-related phenomena (multiple acts of elastic scattering) but also the interference ones (multiple acts of 
consecutive elastic and inelastic scattering). 

For one thing, the former type of the conductivity corrections would depend not only on the elastic scattering rate
but also on some rather special inelastic ones (Cooperon/Diffuson phase-breaking), whereas  
the latter would typically acquire its temperature dependence via 
the fermion occupation factors. Neither source of the $T$-dependence can be readily envisioned in the 
variational approach of Ref.\cite{noMIT}, though. 
While suitable for analyzing classical percolation, this approach operates in terms of the Kirchoff's law
for a random network of classical resistors and employs the  
quadratic Thomson's variational action akin to that for maximum entropy production.
In light of such potential caveats it does not seem surprising that the results of Ref.\cite{noMIT}
suggest the diffusive metallic behavior even at weak repulsive interactions strengths in $2d$ - which 
would then be in conflict even with the perturbative Altshuler-Aronov theory.

Leaving out the remaining possibility of finding a better variational ansatz that could yield a lower conductivity than the 
simplest one of a constant current/density employed in Ref.\cite{noMIT} and resulting in the above lower bound, it is worth noting that the scenario of Ref.\cite{noMIT} might also be limited to the rather special class of models where no coupling between 
the translation symmetry-breaking (Stueckelberg) degrees of freedom and the Maxwell field is allowed.
 
To that end, a further investigation into the issue \cite{nonoMIT} 
showed that, in contrast, no finite (let alone, universal) lower bound seems to exist in the more general models which include 
generic couplings between the axions with the expectation values  
$\Phi_i\sim x_\mu\delta_i^{\mu}$ and vector and/or scalar fields
\be
\Delta L = \sum_{n=1}(g^{\mu\nu}\partial_{\mu}\Phi_i\partial_{\nu}\Phi_i)^n[a_nF_{\mu\nu}^2+b_n\phi^2]
\ee
Thus, conceivably, the scenario of 'many-body delocalization' proposed in Ref.\cite{noMIT} is
neither generic, nor generalizable beyond the scopes of the variational analysis. 
However, the results of Refs.\cite{nonoMIT}, too, should be taken cautiously, 
as, e.g., the second reference finds the conductivity to be potentially 
negative, depending on the strength of the couplings $a_n,b_n$.

In that regard, although these and other predictions would certainly benefit from clear disclaimers 
about the limited range of the holographic parameters for which physically sensible predictions 
(e.g., positive conductivity) can be made, conspicuously,
such limits do not readily follow from the phenomenological 'bottom-up' approach itself. 
In particular, it would be premature to apply any of the aforementioned holographic scenaria
to the analysis of, e.g., the $2DEG$ systems demonstrating the apparent 
metal-insulator transition, complete with its peculiar scaling properties \cite{kravchenko}.\\

{\it Holographic Transport: Dirac/Weyl materials}\\

An important example of the systems whose transport properties
can be studied systematically and then compared to the holographic predictions 
is provided by the interacting Dirac/Weyl (semi)metals.
In the past, the transport properties of (pseudo)relativistic systems have already been addressed in the 
context of, both, interacting $2d$ bosons  and fermions \cite{damle}.
Such studies utilized the standard method of quantum kinetic equation
which offers a viable insight that can be compared to the specific predictions of the 
general theory of holographic transport. 
The fermionic variant of this problem pertinent to the case of Coulomb-interacting graphene
was studied in the so-called two-mode approximation where 
the emphasis was made on the conservation of charge and energy densities, 
as well as momentum density which is equivalent to the energy current up to the higher-order dissipative corrections
(which feature is specific to the (pseudo)relativistic dispersion)  \cite{graphene}.

In the case of a weak Coulomb coupling $\alpha=e^2/hv<<1$
the equilibration rate of all other, non-conserved, modes would be of order 
$\Gamma_{non-cons} \sim \alpha^2|\ln\alpha |T$ which is higher than 
that of the conserved ones, $\Gamma_{cons}\sim\alpha^2T$, 
thanks to the $2d$ kinematic (logarithmic) divergence of the Coulomb collision integral.
Thus, the conserved modes can be singled out, while neglecting all those that undergo a faster relaxation \cite{graphene}.

It turns out, however, that the system possesses another (nearly)conserved density
which is the 'imbalance' mode corresponding to the total number of electrons and holes 
(as opposed to their difference related to the charge density) \cite{karlsruhe}. 
Away from the neutral (particle-hole symmetric) regime  
its relaxation takes place through the Auger-type processes with the still slower rate of order $\Gamma_{imb}
\sim T^4/\mu^3$.
Close to the neutrality point the results of the two- and three-mode analyses
appear to differ somewhat, thus showing that the imbalance mode 
can indeed impact the low-energy hydrodynamic behavior \cite{karlsruhe}.

It would, therefore, be desirable to carry out the three-mode calculations for the 
$SU(N)$-symmetric fermions with $N\gg 1$, focusing on the regime $1/N\ll\alpha^2\ll 1$
where a comparison with the holographic predictions can be made \cite{future}.

Notably, the kinetic equation tends to yield the conductivity which 
develops the Drude peak followed by a dip at $\omega\sim \Gamma$,
while no other peaks (nor zeros) emerge,
in contrast to the holographic/hydrodynamic/memory matrix result (30).
Besides, it remains to be seen as to how the 
explicitly computed corrections to the optical conductivity of graphene \cite{mishchenko}
\be
{\sigma(\omega)\over \sigma(\infty)}=1+{C\over {|\ln\omega|}}
\ee
match with the general prediction \cite{krempa_last}
\be
{\sigma(\omega)\over \sigma(\infty)}=1+\sum_n{C_n({T\over \omega})^{\Delta_n}}
\ee
where the sum is taken over the operator expansion of the product of two current operators.

Another quantity of special interest would be the Lorentz ratio (ordinary, as well as the Hall one) 
as a function of frequency. In the $DC$ limit and near the neutrality point 
this ratio is dramatically enhanced because the heat and charge currents are controlled 
by two different mechanisms: momentum relaxation vs electron-hole 
pair creation. This prediction was found to be in a good agreement with the recent data on graphene \cite{kim}
showing a $20$-fold enhancement
(although the ratio $\eta/s\sim 10$ estimated in Ref.\cite{kim} seems to 
indicate that even a free-standing graphene may not be $\bf that$ strongly coupled, after all). 

It would also be interesting to generalize the analysis to $d=2+\epsilon$ dimensions
where the effects of the Coulomb interactions are expected to be less dramatic and the separation 
between the slowly vs rapidly decaying modes becomes less pronounced \cite{future}. 

Besides, one would need to introduce yet another nearly conserved density describing an 
imbalance between the numbers of quasiparticles in the vicinity of the two different Weyl points. 
This 'chiral' density incorporating (for $\epsilon=1$) the effects of the $3d$ chiral anomaly 
gives rise to such concomitant transport phenomena as a negative magnetoresistance \cite{Weyl}.
Some of its underlying geometric aspects can be observed even at
the semiclassical level where the phase-space dynamics is naturally
described in terms of the Bloch curvature \cite{bloch}
$
\Omega_k=\epsilon_{ij}<{\partial\Psi\over \partial k_i}|{\partial\Psi\over \partial k_j}>
$
which affects the semiclassical equations of motion
\bea
{d{\bf r}\over dt}={\partial \epsilon\over \partial {\bf k}}-{\bf \Omega_k}\times {d{\bf k}\over dt},\nonumber\\
{d{\bf k}\over dt}=-{\partial \epsilon\over \partial {\bf r}}+{\bf \Omega_B}\times {d{\bf r}\over dt}
\eea
The current relaxation is then described by the equations
\bea
\partial_{\nu} T^{i\nu}=F^{i\nu}J_{\nu}-\gamma T^{it},
\nonumber\\
\partial_{\nu}J^{\nu}=\alpha \epsilon_{\mu\nu\lambda\rho}F^{\mu\nu}F^{\lambda\rho}
\eea
Reproducing these effects in their entirety would provide another check point 
for (and further advance) the holographic transport theory \cite{sun}.
 
In that regard, one should also mention a series of works aimed at connecting holography to the more conventional 
field theoretical analysis \cite{stoof}. It seems, though, that the actual graphene and $3d$ 
Weyl/Dirac metals alike can hardly provide a viable playground for the CMT holography, 
as the effects of interactions in these materials are either mild or even outright weak.  
It would, therefore, be quite interesting to find some realization of their counterparts 'on steroids'
in the condensed matter realm.

The extended multi-mode kinetic analysis of graphene may also allow for a deeper understanding of the
role of viscosity in electron transport. The viscous terms 
become relevant for $\eta>(enw)^2/\sigma$ where $w$ is the width
of the graphene sample and such analysis can
even be performed in the framework of the Navier-Stokes equation, this time in the real space \cite{levitov}. 
Incidentally, this regime has recently been studied in the ultra-pure $2d$ metal $PdCoO_2$ \cite{PdCoO},
although the estimated ratio $\eta/{\cal S}\sim 10^6$ can hardly allow one to view it as ideal fluid. 

Also, the problem of a hot spot relaxation in graphene \cite{karlsruhe}
presents a specific example of the general 'quench'-type phenomena 
which, too, have been extensively addressed by the CMT holography \cite{quench}.
The $1d$ variant of this problem involves two reservoirs 
of different temperatures $T_{L}$ and $T_{R}$ which are brought into a thermal contact,
its solution featuring a non-equilibrium  stationary state 
characterized by a definite temperature $(T_{L}T_R)^{1/2}$  
and constant energy flux (this solution does not seem to readily extend to the 
limit of either vanishing temperature, though). Moreover, this stationary state 
was shown to spread outward in the form of shock waves
and was related to the Lorentz-boosted black-hole metric in the bulk, 
thus suggesting yet another explicit holographic connection.

To that end, it was conjectured in  Refs.\cite{quench} that a similar behavior would occur in higher dimensions as well.  
However, the recent analyses of the hot spot relaxation in $2d$ graphene \cite{karlsruhe} suggest 
that a more involved scenario  where, both, the resistive and viscous effects 
interfere with one another might be realized. Notably,  
the recent holographic work of Refs.\cite{quench_last}, too, shows a more involved 
dynamics of the expanding stationary state.\\

{\it {Analogue Holography Demonstrators}}\\

The above examples of 'emergent geometry' prompt a systematic 
quest into the apparent holography-like relationships that would be governed by the already known, rather than some hypothetical, physics.
To that end, the general holographic concept can benefit from the possibility of being simulated 
in various controlled 'analogue' environments.

One prospective design of a 'holography simulator' 
was proposed for implementation in flexible graphene and other semi-metallic monolayers (silicene, germanine, 
stanine, etc.) \cite{btz}.
Such stress-engineered desktop realizations of the system of $2d$ Dirac fermions in a curved geometry  \cite{iorio} 
can also be grown on commensurate substrates (e.g., $h-BN$) in order to endow the bulk fermions 
with a finite mass via hybridization.

In Ref.\cite{btz} a number of situations were discussed   
where a physical edge of the curved graphene flake which supports (almost)
non-interacting massive Dirac fermions propagating in a curved $2d$ space exhibits a behavior that would be typically attributed to the effects of some phantom $1d$ interactions. In contrast to the aforementioned 'semi-local' 
scenario of Refs.\cite{semi}, though, it is the momentum 
dependence of the $1d$ propagator $G(\omega,k)$ that tends to become non-trivial.

Specifically, in graphene the artificial gauge and metric fields represent the elastic (phonon) degrees of freedom, 
their effective vector and scalar potentials 
\be 
A_0\sim u_{xx}+u_{yy},~~A_x\sim u_{xx}-u_{yy},~~A_y\sim u_{xy}
\ee
 being 
composed of the components of the strain tensor \cite{vozmediano},  
$
u_{ij}={1\over 2}(\partial_iu_j+\partial_ju_i+\partial_ih\partial_jh)
$.

Here $u_i({\bf x})$ and $h({\bf x})$ are the local in- and out-of-plane displacements of the monolayer,
respectively, while the valley-specific fields $A_\mu$ have opposite signs at the two different Dirac points.
Computing the massive free Dirac fermion propagator in a curved space 
and taking its boundary limit one obtains the large-scale asymptotic behavior 
(4) governed by the geodesic action $S_{hol}(\tau,x)$ for the background metric.
For instance, in the case of a graphene flake shaped as a surface of rotation 
with the line element (here $r$ and $\phi$ are the normal polar coordinates) 
\be
dl^2=dr^2[1+({\partial h(r)\over \partial r})^2]+r^2d\phi^2
\ee
with $h(r)\sim (R/r)^\eta$ for $r\geq a$ the 'warp factor' $g(r)\sim 1/r^{2\eta+2}$
diverges at small $r$. One then obtains the geodesic action 
\be
S(\tau,x)=m{\sqrt {\tau^2+(Rx^{\eta})^{2/(\eta+1)}}}
\ee
which reveals an unconventional behavior of the boundary propagator as a function of the distance along the edge,
thus suggesting the 'holographic' dynamical critical exponent $z={\eta/(\eta+1)}$.

It is instructive to compare the asymptotic (4) with the action given by (44) with the propagator of $1d$ fermions interacting via a pairwise potential $U(x)\sim 1/x^\zeta$ where $\zeta<1$. Using the standard $1d$ 
bozonization technique and matching the large-$x$ asymptotics, one finds that (44)
can mimic the spatial decay of the $1d$ propagator in the presence of such interactions, provided that
 $\eta=(1-\zeta)/(1+\zeta)$. 

Another instructive example is provided by the line element  
\be
dl^2_{log}=dr^2+R^2\exp({-2(r/R)^\lambda})d\phi^2
\ee
in which case the geodesic action reads
\be
S(\tau,x)=m{\sqrt {\tau^2+R^2(\ln x/a)^{2/\lambda}}}
\ee
For $\lambda=1$ and at large $x$ the propagator decays algebraically, $G(0,x)\sim 1/x^{mR}$ 
which is reminiscent of the behavior found in the $1d$ Luttinger liquids,
while for $\lambda\neq 1$ Eq.(46) yields a variety of stretched/compressed exponential  
asymptotics which decay faster (for $\lambda<1$) or slower (for $\lambda>1$)
than any power-law. For instance, by choosing $\lambda=2/3$ one can simulate a 
faster-than-algebraic spatial decay, $G(0,x)\sim \exp(-const\ln^{3/2}x)$, in 
the $1d$ Coulomb gas ($\sigma=1$) which is indicative of the incipient $1d$ charge density wave state.

For other values of $\lambda$ Eq.(46) reproduces the behavior in the boundary theory governed by the interaction $U(x)\sim (\ln x)^{(2/\lambda)-3}/x$. Although the physical origin of such a bare potential would not be immediately clear, multiplicative logarithmic factors do routinely emerge in those effective $1d$ interactions ('double trace operators')
that are associated with various marginally (ir)relevant two-point operators.
On the experimental side, the boundary correlations can be probed with such established techniques as time-of-flight, edge tunneling, and local capacitance measurements, thus potentially 
helping one to hone the proper analogue-holographic 'dictionary'. 

This holography-like correspondence once again suggests that some limited form of a bulk-boundary relationship might, 
in fact, be quite robust and hold regardless of whether or not the systems in question possess any particular 
symmetries, unlike in the original $AdS/CFT$ construction.
Naively, this form of correspondence can even be related to the Einstein's equivalence principle
(i.e., 'curvature equals force'), according to which free 
motion in a curved space should be indistinguishable from the effect of a physical interaction (only, this time around, 
in the tangential direction). 

As an alternate platform for doing analogue holography, optical metamaterials have long been 
considered as candidates for simulating such effects of general relativity
as event horizons, redshift, black, white, and worm holes, inflation, 
Hawking radiation,  dark energy, multiverse, Big Bang and Rip, metric signatures transitions, 'end-of-time', 
and other cosmological scenaria.

However, the earlier proposals focused on the effective $3+1$-dimensional 
metrics which requires some intricate engineering of the locked permittivity ($\epsilon$)  
and permeability ($\mu$) tensors \cite{meta1}
\be
\epsilon^{ij}=\mu^{ij}=g^{ij}{{\sqrt -g}/|g_{00}|}
\ee 
More recently, it was proposed to use extraordinary (TM-polarized) monochromatic photons 
with the dispersion $\omega^2=k_z^2/\epsilon_{xy}+k^2_{xy}/\epsilon_{zz}$
for simulating the effective $2+1$-dimensional metrics  
\be
g_{\tau\tau}=-\epsilon_{xy}, ~~~g_{rr}=g_{\phi\phi}/r^2=-\epsilon_{zz}
\ee
in the media with $\mu=1$  \cite{meta2}.

In the hyperbolic regime, $\epsilon_{xy}>0$ and $\epsilon_{zz}<0$, the momentum component $k_z$
can then be thought of as an effective frequency, whereas $\omega$ plays the role of mass.
In Ref.\cite{meta} it was shown that the experimentally attainable two-component metamaterial configuration consisting of
alternating metallic and insulating (flat or cylindrical) layers can simulate some of the HV geometries.  
The boundary propagator describing static spatial correlations of the optical field 
on the interface between the metamaterial and vacuum was found to behave as 
\be
<E_{\omega}({\bf x})E_{-\omega}(0)> \sim\exp[-|\omega{\bf x}|^{\theta/z}]
\ee
which should be contrasted to its counterpart in an isotropic medium with a (negative) dielectric constant,  
$
<E_{\omega}({\bf x})E_{-\omega}(0)>\sim \exp(-\omega |{\bf x}|)
$.
Experimentally, such correlations can be studied by analyzing the 
statistics of a non-local optical field distribution with the use of holographic and 
speckle interferometry. Albeit still requiring a careful engineering of the dielectric media, some hallmark features of the analogue holography could, in principle, be detected 
and then compared with, e.g., correlators of the vertex operators in the theory of a
strongly self-interacting $2d$ bosonic field, akin to that describing the thermodynamics of
a fluctuating elastic membrane. 

Lastly, one could further elaborate on yet another (historically, far more extensively studied) potential
playground for analogue holography which is provided by the acoustic realizations of 'emergent gravity' \cite{visser}.
By implementing such proposals 
one might also hope to establish a putative bulk-boundary correspondence (alongside its own 'dictionary') for a broader variety of
the combined (pseudo)gravitational backgrounds, thereby gaining a better insight into the physical origin(s) of
the apparent holography-like properties of various CMT and AMO systems. 

Together with the systematic comparison 
between the predictions of the CMT holography and other, more traditional, approaches and/or experimental data
it would be a necessary step towards vetting the intriguing, yet speculative, holographic ideas 
before the latter can be rightfully called  a novel  
technique for tackling the 'strange metals' and other strongly correlated materials. 
In any event, though, it would seem rather unlikely that a hands-on expertise in string 
theory will become a mandatory prerequisite for understanding the cuprates. 

The author gratefully acknowledges the support from and/or hospitality at the Aspen Center for Physics
(funded by the NSF under Grant 1066293), the Galileo Galilei Institute (Florence),
the Kavli Institute for the Physics and Mathematics of the Universe (Kashiwa),
the Kavli Institute for Theoretical Physics (Beijing), the International 
Institute of Physics (Natal), and NORDITA (Stockholm) during the various stages of compiling these notes.

\end{document}